\documentclass[%
 aip,
 amsmath,amssymb,
 reprint,%
]{revtex4-1}
\usepackage[textsize=tiny]{todonotes}

\usepackage{graphicx} % Required for inserting images
\usepackage{braket}
\usepackage[utf8]{inputenc}
\usepackage{textgreek}

\begin{document}
\title{The seeds of the future are in the present: A blind exploration of metastable states}
\author{Timothée Devergne}
 \affiliation{Atomistic Simulations, Italian Institute of Technology, Genova, Italy}%Lines break automatically or can be forced with \\
 \affiliation{Computational Statistics and Machine Learning, Italian Institute of Technology, Genova, Italy}
\author{Vladimir Kostic}%
\affiliation{%
Computational Statistics and Machine Learning, Italian Institute of Technology, Genova, Italy%\\This line break forced% with \\
}%
\affiliation{Department of Mathematics and Informatics, University of Novi Sad, Novi Sad, Serbia}

\author{Massimiliano Pontil}
\affiliation{%
Computational Statistics and Machine Learning, Italian Institute of Technology, Genova, Italy%\\This line break forced% with \\
}%
\affiliation{ AI Centre, Department of Computer Science, UCL, London, UK}
\author{Michele Parrinello}
\affiliation{Atomistic Simulations, Italian Institute of Technology, Genova, Italy}
\begin{abstract}
    In this work, we present a novel type of molecular dynamics simulation that aims at discovering, in a blind way, new metastable states. Using only data coming from an initial unbiased simulation, and with the help an appropriately defined loss function, we compute a bias that favors sampling yet unexplored configurational space regions, encouraging the system to leave the initial basin. In our work, we take advantage of what is normally thought to be a defect, namely the difficulty of neural networks to generalize. Contrary to most other enhanced sampling methods, which need previous knowledge of the reactive process, we are able to discover in a blind way new metastable states, overcoming otherwise insuperable kinetic bottlenecks. We illustrate the workings of the method with a number of instructive examples.
\end{abstract}
\maketitle

\section{Introduction}
\begin{figure}
    \centering
    \includegraphics[width=0.9\linewidth]{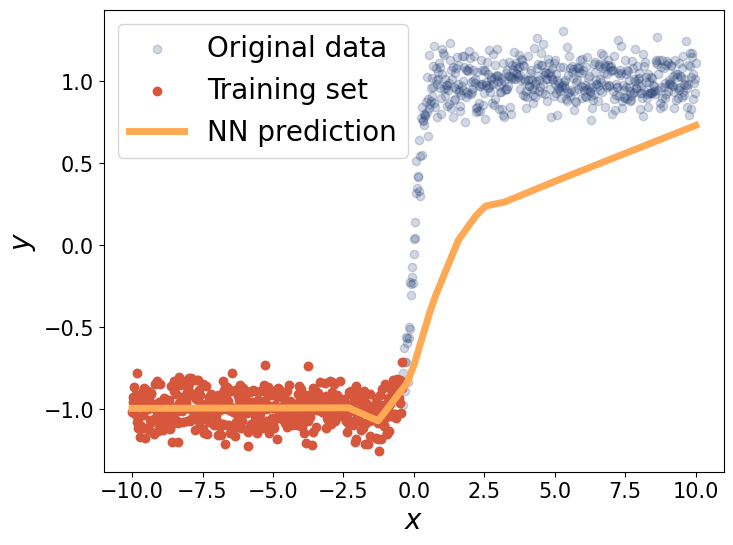}
    \caption{Example of extrapolation for an objective data following a hyperbolic tangent function. A neural network model was trained only on a part where the target function is constant. The neural network extrapolation behavior is illustrated for $x>0$ }
    \label{fig:tanh}
\end{figure}

Molecular dynamics simulations based on an atomistic description of matter play an increasingly relevant role in modern science.  Such simulations help understanding and guiding experiments, and finally, gaining insight into physical phenomena. They can also replace experiments and act as a virtual microscope of very high resolution. In most areas of physics and chemistry, it is rare to find experimental papers that are not accompanied by an atomistic simulation \cite{sponer_prebiotic_2016,perez-villa_synthesis_2018,meuzelaar_folding_2013}. 

One appealing feature of molecular dynamics simulations is that, at least in principle, they proceed like laboratory experiments. One sets up the numerical experiment by choosing a potential that describes as accurately as possible the interatomic interactions and also the thermodynamic conditions of the experiment. Then, lastly, and most importantly, once enough data from the simulation has been collected, one tries to understand the science behind the results obtained.  

With the help of machine learning, the problem of designing accurate interatomic potentials has made important strides forward\cite{perego_data_2024,behler_generalized_2007,unke_machine_2021,behler_four_2021}. Ways of controlling the thermodynamic conditions have also been developed \cite{nose_unified_1984,hoover_canonical_1985, bussi_canonical_2007}.   However, still severe technical problems remain when simulating systems which evolve by visiting different metastable states separated by large barriers. Examples of such scenarios are first-order phase transitions like crystallization, most chemical reactions, or drug-protein binding. The kinetic bottlenecks associated with such phenomena make observing transitions between long-lived metastable states extremely unlikely\cite{pietrucci_strategies_2017}.   These transitions are often referred to as rare events. Their study has been the subject of a large number of investigations \cite{laio_escaping_2002,bolhuis_transition_2002,invernizzi_rethinking_2020, invernizzi_exploration_2022} and, although great progress has been made,  much remains to be done.

Most enhanced sampling methods assume that the initial ($A$) and final  ($B$) metastable state are known beforehand \cite{france-lanord_data-driven_2024,bonati_data-driven_2020,trizio_enhanced_2021}, either explicitly as in the case of path based methods \cite{bolhuis_transition_2002} or implicitly, as in the case of metadynamics \cite{laio_escaping_2002}  where the choice of collective variables (CVs) encodes also the $B$ state.
  Despite the many successes, there is something practically and intellectually unsatisfactory in restricting oneself to this scenario. Often, the $B$ state is not known beforehand, and having a method capable of discovering new metastable states would be of the greatest importance.  The present state of affairs also implies that molecular dynamics is not yet that perfect method which, like in an experiment, interrogates the system in an unbiased way and comes out with the result.

Here we propose a method that makes this type of simulations possible, allowing the system  overcoming  kinetic barriers and moving  from metastable state to metastable state without any previous knowledge of the system. Several attempts have been made already to achieve this result.  Metadynamics-like approaches \cite{pietrucci_graph_2011, raucci_discover_2022} have suggested using as CVs the eigenvalues of the adjacency matrix, which encode the molecular structure \cite{ismail_graph-driven_2022}.  With this method, successful predictions of reactions relevant to atmospheric \cite{yang_molecular_2025} and high-pressure chemistry \cite{thevenet_water_2025} have been reported. However, this approach cannot be easily extended beyond the realm of atmospheric chemistry. More recently, Zhang and Piccini \cite{zhang_exploring_2025} in an elegant approach have suggested using the skewness of the $A$ distribution to drive the system out of the initial metastable state.

Here we take an approach similar in spirit to that of Zhang and Piccini \cite{zhang_exploring_2025}, but one that is more general since it does not rely on quasi-harmonic considerations and which leads to highly non-local bias that can act far away from $A$. Our method is based on a novel 
%new and more general 
principle that might have general relevance in machine learning and allows   automatically discovering several metastable states in sequence.

We use here methodologies developed in our recent studies of the committor  \cite{kang_computing_2024, trizio_everything_2025} and of the closely related dynamics operator \cite{devergne_biased_2024, devergne_slow_2025}.  Following these papers we introduce a loss function that, when minimized using only data in A, has as solution a function $I(x)$ that is constant  (i.e. $I(x)\approx 1$) in the sampled region of $A$, but deviates from being a constant outside, where it has to extrapolate. It has been argued that, in this regime, the neural network extrapolates linearly \cite{xu_how_2020}, but for what concerns us here, it only matters that the derivative of $I(x)$ changes its value when going outside the already sampled region. The function  $\vert \nabla I(x)\vert^2$  is thus approximately $0$ in the convex hull of  $A$ and changes to a finite value outside. This is illustrated in FIG. \ref{fig:tanh}, where we generated data following a hyperbolic tangent, which is a simplified model for a committor. We then trained a model using only samples coming from the left of the step. It can be seen that the slope of the model clearly changes when we leave the training set.

\begin{figure}[htbp]
    \centering
    \includegraphics[scale=0.3]{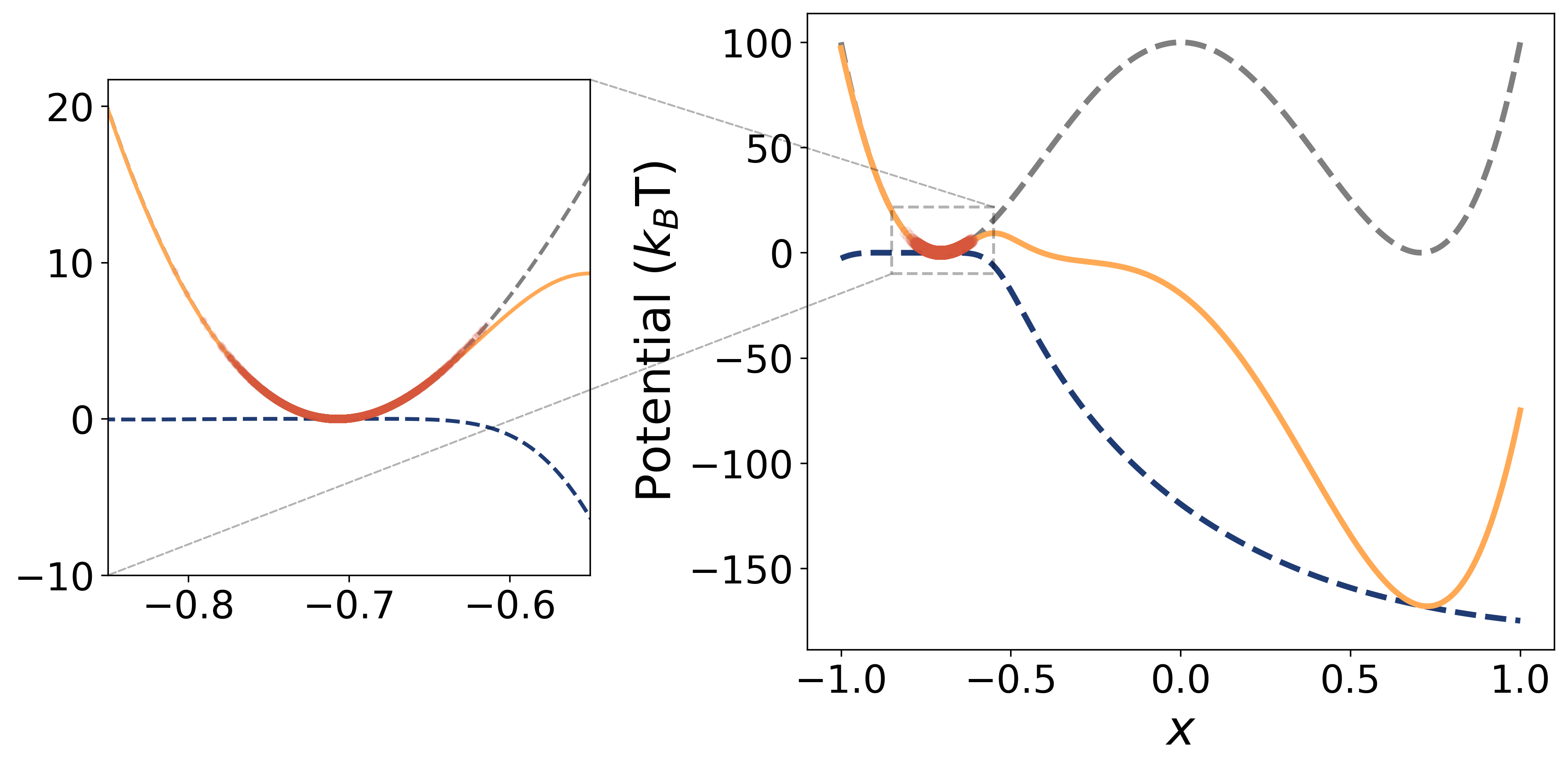}
    \caption{Kolmogorov bias from a model trained with data only from the left state of a double well potential. The dashed gray lines indicate the underlying potential, the dashed blue lines the Kolmogorov bias. The orange line is the sum of the Kolmogorov bias and the potential. Finally, the red points are the training data.}
    \label{fig:MB}
\end{figure}

To attract the system towards regions of the configurational space where $\vert \nabla I(x)\vert^2$ is non-zero zero, we can use the following potential by taking inspiration from our works on the committor function \cite{kang_computing_2024,trizio_everything_2025}
\begin{equation}
    V_\mathcal{K}(x) = -\frac{\lambda}{\beta} \log{(|\nabla I(x)|^2+ \epsilon)} +  \frac{\lambda}{\beta} \log{\epsilon} 
\end{equation}
where $\lambda$  modulates the bias strength and $\epsilon$ is a regularization parameter. This will tend to push sampling towards the edges of the already sampled  $A$ region.  We also note that the maximum possible value of $V_{\mathcal K}(x)$  is $0$, hence outside the already explored region it has to be negative.  Given its link to the functional introduced by Kolmogorov \cite{kolmogoroff_uber_1931}, we shall refer to $V_{\mathcal K}(x)$ as the Kolmogorov bias.
The way the Kolmogorov bias acts is illustrated in figure \ref{fig:MB}: where we show the behavior of $ V_\mathcal{K}(x)$ in a double well potential, when $I (x)$ is trained on left basin data only. It can be seen that the Kolmogorov bias removes the barrier at the right of A and even lowers the B state. We also note that the extrapolation reflects the skewness of the potential, and as discussed by Zhang and Piccini \cite{zhang_exploring_2025}, this will help moving to $B$. In addition, since it is defined for any $x$ it will make itself felt all over the space, adding a field that encourages the system to visit yet unexplored regions.

\section{Results}
\begin{figure}
    \centering
    \includegraphics[scale=0.22]{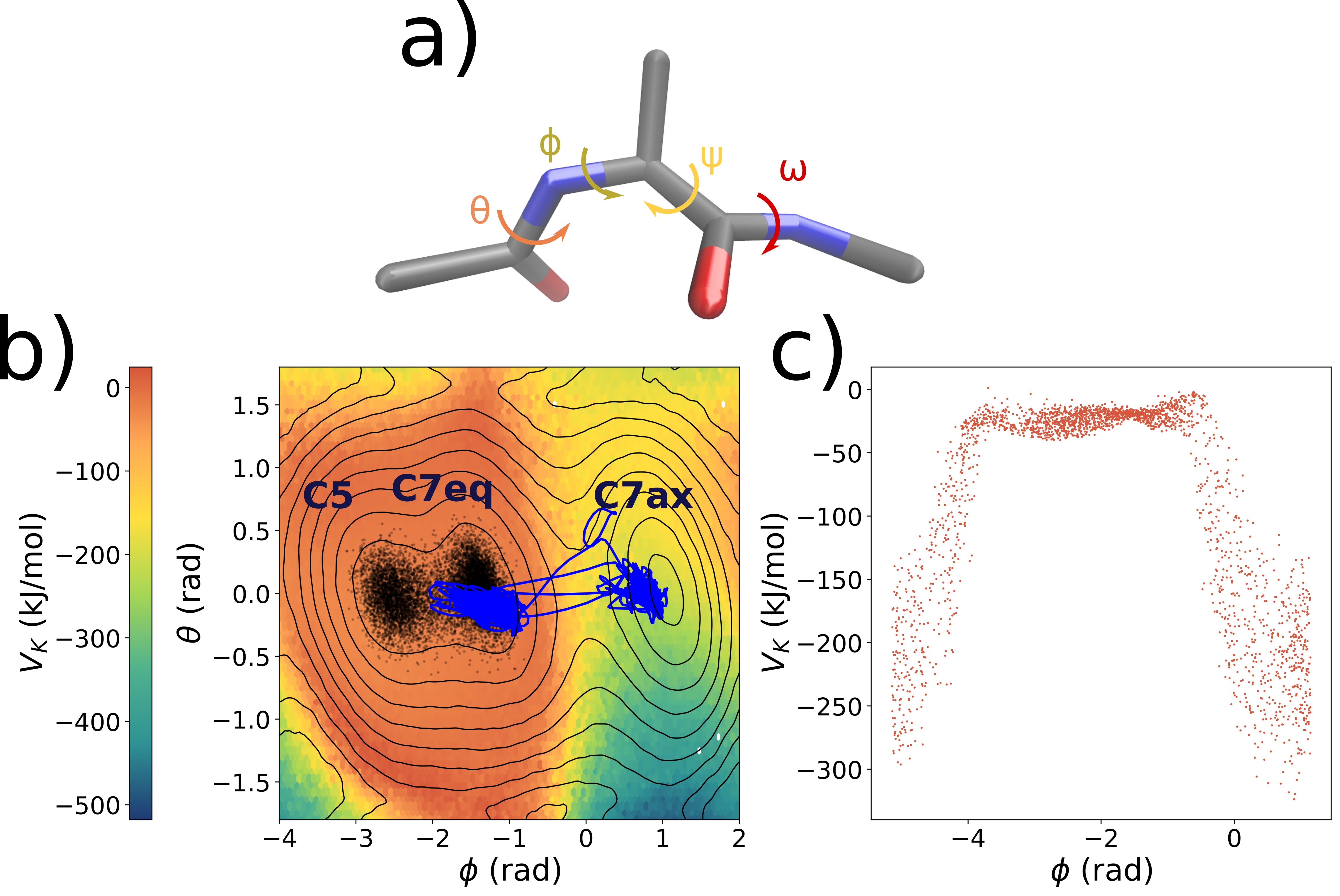}
    \caption{Panel a):  Alanine dipeptide and its four dihedral angles. 
     %It can be seen that in principle four metastable states are possible. The lowest ones $C5, C7_{eq},\,\text{and}\, C7_{ax}$ are  located in the upper part of the surface 
    b): (blue) An example of an alanine dipeptide trajectory under the combined action of the Kolmogorov bias and of \textsc{OPES} explore metadynamics for a model that uses the dihedral angles as descriptors. %Panel c): (blue)  Same as b)  using  the interatomic distances as descriptors. Panel d): same setup as panel c), but with  a cutoff  added to the Kolmogorov bias. In all panels, the training points are represented as black dots. 
  The isolines of the underlying free energy surface are drawn, and the background color represents the Kolmogorov bias.
    c): values of the Kolmogorov bias along the $\phi$ angle, as expected, it has a mesa-like shape, and it is close to zero in the sampled region.}
    \label{fig:alanine}
\end{figure}

In this section, we report the results obtained in order of complexity of the system studied. We first start with the sempiternal example of alanine dipeptide in vacuum to show that from an unbiased simulation in one metastable state, we can escape this state and move to another conformation. Then, we show that we can explore chemical reactions, with the example of a Claisen rearrangement \cite{noauthor_claisen_nodate,krenske_claisen_2017} in the molecule of 3-ethenoxyprop-1-ene which has already been used as a test for other reaction discovery schemes \cite{raucci_discover_2022}. Finally, we show that even for higher-dimensional systems, our method can discover new states in multiple-state systems by tackling the slightly more complex molecule, alanine tetrapeptide.

\subsection{ Alanine Dipeptide}

Alanine dipeptide is a simple molecule which presents conformational changes associated with the flexibility of its dihedral angles. A very large body of experience has been accumulated on this molecule\cite{bonati_data-driven_2020,branduardi_b_2007, bonati_deep_2021,frohlking_deep_2024}. Its conformational space is spanned by the four dihedral  angles   $\phi\, ,\psi \, , \theta \,$ and $\omega$ (see FIG. \ref{fig:alanine}, panel a)) . The angle  $\phi\,$ is known to be a good CV, and $\theta$ has been found to be part of the reaction coordinate \cite{kang_computing_2024, bolhuis_reaction_2000}.  

The lowest part of the free energy surface projected on the  $\phi\, ,\theta$  plane exhibits three well-known conformational states: C5, C7eq, and C7ax. A low barrier separates the C5 and C7eq minima, and at the simulation temperature (T=300K), these conformers can easily interconvert. The barrier to go from these two low-lying states to  C7ax  is much higher and needs some kind of enhanced sampling method to be overcome.
The lowest metastable state is thus multimodal, being a combination of the  C5 and  C7eq conformers. In one such case, the skewness of the distribution does not necessarily identify the escape direction.

In this system, it is natural to use the four dihedral angles introduced above as descriptors.  The resulting value $V_\mathcal{K}(x)$ projected on the $\phi, \theta$  plane has the desired properties of being strongly attractive towards  C7ax.  Since obtaining  $V_\mathcal{K}(x)$  is the result of a stochastic process, different models can extrapolate differently.  However, while in all circumstances   $V_\mathcal{K}(x)$   always encourages the system to exit $A$, different models can do so with different efficiency.  We will therefore compare the performance of our approach by studying an ensemble of models. Thus, we trained 50 different models with different initial random parameters on the same data.  For each model, we ran a 5 ns long simulation. We set the \textsc{opes barrier} parameter to 40kJ/mol,  as the expected barrier to cross to    C7ax  is of this order of magnitude. Of the  50 simulations, 39 went to the C7ax basin (see FIG. \ref{fig:alanine} panel b) for one example), while 2 visited a higher energy state (see SI). Finally, to better control the behavior of the system, one can add a cutoff to the Kolmogorov bias to avoid exploring high free energy regions (see SI). In FIG. \ref{fig:alanine}, panel c), we show $V_{\mathcal{K}}(x)$ as a function of the $\phi$ angle. Similarly to the double well potential example (see methods section),  $V_{\mathcal{K}}(x)$  is constant in the training basin. In addition it makes the energy barrier between the two wells vanish and lowers the second basin. The probability of exiting $A$ can be further enhanced if one uses multiple walkers\cite{raiteri_efficient_2006} or if appropriate, a multithermal sampling\cite{invernizzi_unified_2020} .

While the choice of the dihedral angles as descriptors is natural, one might suspect that observing these transitions was made possible by our physical understanding of the system.   For this reason, we play ignorant and use as descriptors the  45 interatomic distances between heavy atoms.  In such a case, things are slightly more difficult,  since this set of descriptors describes the angular arrangement only implicitly. We report in the SI such simulations, and of course, the efficiency is reduced, but still a large number of transitions to the $B$ state could be observed. Interestingly, several transitions to conformational states that are higher in energy were observed.

\subsection{An example of chemical reaction discovery}

\begin{figure}
    \centering
    \includegraphics[scale=0.3]{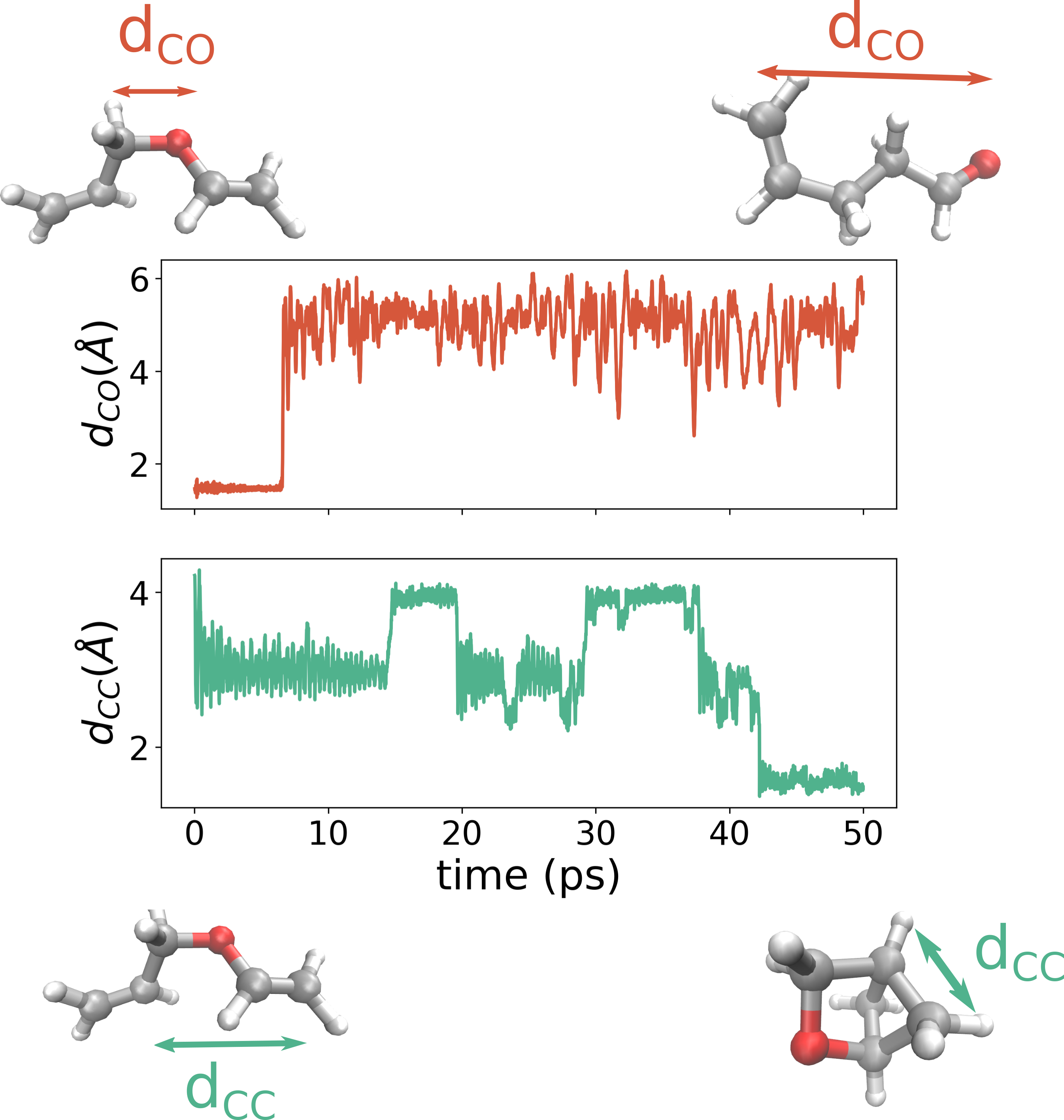}
    \caption{Results of our simulations for the Claisen rearrangement of the  3-ethenoxyprop-1-ene molecule. Top panel: example of a successful Claisen rearrangement.Bottom panel: }
    \label{fig:clasien}
\end{figure}
One example in which this method proved extremely useful was the case of chemical reactions. The example studied here was the reactive behavior of 3-ethenoxyprop-1-ene. Our aim once again was to use as little chemical intuition as possible, and thus,  instead of using an intuitive CV, we used again as descriptors the interatomic distances between all heavy atoms of the molecule. One of the products is the result of a Claisen rearrangement, as shown in ref. \citenum{raucci_discover_2022} leading to the pent-4-enal as expected. To explore the possible pathways, we trained ten models with different initial training parameters, and performed simulations with a \textsc{barrier} parameter of 200kJ/mol.
To prevent the system from exploring regions with a too high free energy, we put a cutoff to the value of the Kolmogorov bias at 400kJ/mol.
In these simulations, we discovered two different new states: the expected pent-4-enal 5 (FIG. \ref{fig:clasien} top panel) and the higher in energy   2-oxabicyclo[2.1.1]hexane state(FIG. \ref{fig:clasien} bottom panel), which was also reported in ref. \citenum{raucci_discover_2022}. These numerical simulations show that using this method, it is possible to explore efficiently the conformational space and discover possible chemical reactions in a blind and efficient way. 

\subsection{Exploring a multi-state system: alanine tetrapeptide}

\begin{figure}
    \centering
    \includegraphics[scale=0.4]{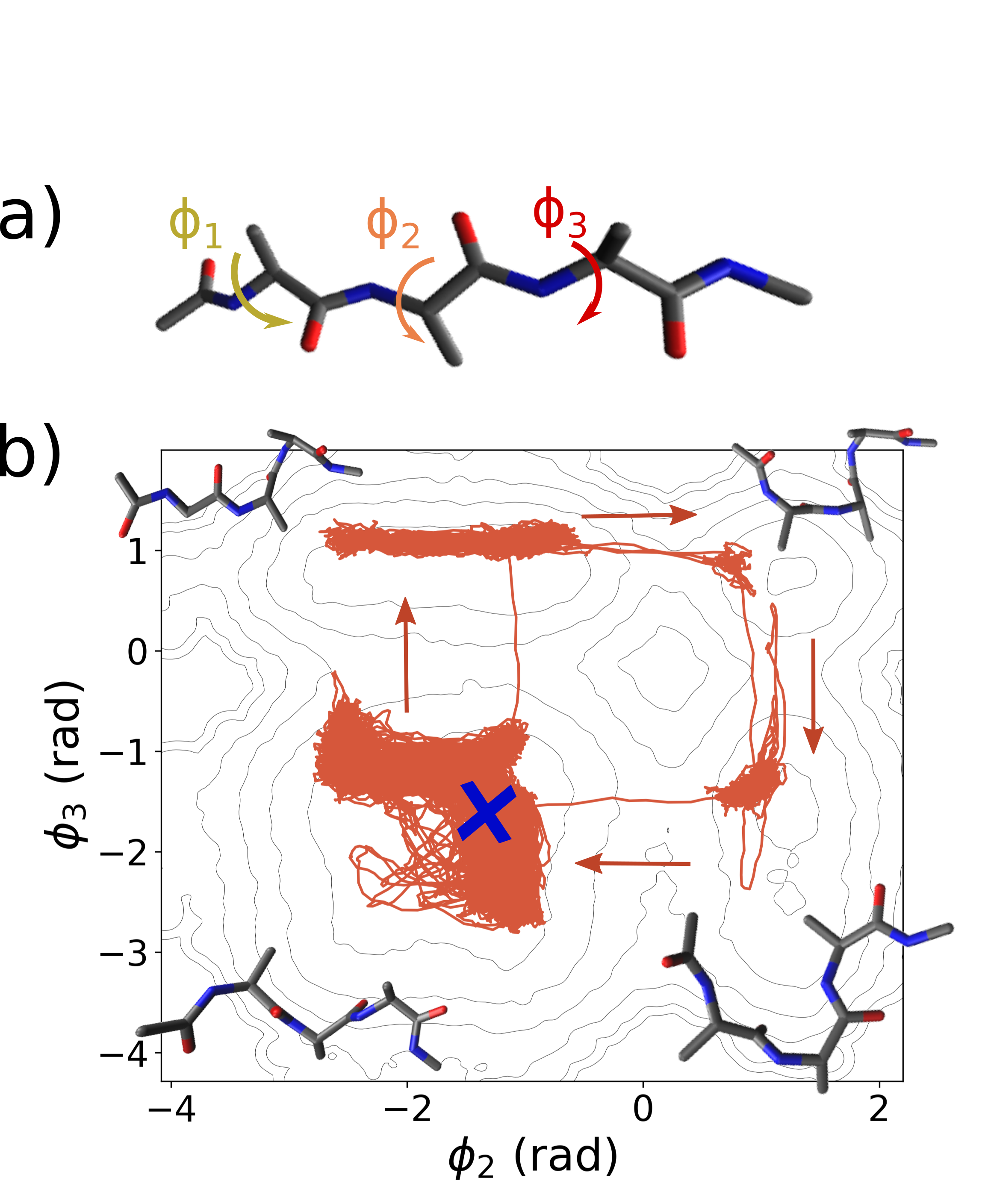}
    \caption{Result of one exploration simulation for the alanine tetrapeptide, in the $(\phi_2, \phi_3)$ plane. The starting basin is indicated by a blue cross.}
    \label{fig:tetrapeptide}
\end{figure}

We now apply our exploration method to the conformational changes of a molecule that displays more than two accessible metastable states: alanine tetrapeptide. The natural variables of the problem are the three dihedral angles ($\phi_1, \phi_2, \phi_3$) \cite{hovan_defining_2019, tsai_sgoop-d_2021} (see FIG. \ref{fig:tetrapeptide}), panel a) ). In this representation, the four slowest transitions are related to changes in the sign of the  $\phi_2$ and $\phi_3$  angles, for a total of four metastable basins.  Once again, in order to be as blind as possible in the exploration phase, instead of using the dihedral angles, which are natural CVs, we used as input of the neural networks all the 190 interatomic distances between heavy atoms, illustrating the ability of our method to deal with high-dimensional descriptor spaces. We show in FIG. \ref{fig:tetrapeptide} the results of one such simulation that visited all 4 basins in the allotted 20 ns simulation time. This happens because of the long-range nature of $V_{\mathcal{K}}(x)$, which also acts far away from A and encourages the system to visit other metastable states. This illustrates the ability of our method to explore several unknown basins, as in the case of alanine dipeptide. In the supporting information, we show the results for other models.

\section{Discussion and Conclusion}

Our method allows a blind discovery of new metastable states based only on data coming from one state and does not requires the use of a CV. However, the choice of relevant descriptors might ease the task, as we have seen in the case of alanine dipeptide.%, where it was easier to exit the basins using dihedral angles as descriptors rather than distances. 

%In this case, the angles are direct descriptors of the transition and, therefore, fluctuations in these variables, as enhanced by our approach, favor the transitions. In contrast, distances are only indirectly related to the reaction coordinates, and enhancing their fluctuations will result with greater difficulty in a transition.  It is, however, indicative of the robustness of the    method that the relevant transitions have been detected, even when using  poor descriptors. 
Methods for selecting the most appropriate descriptors can be devised,  either by using when appropriate what is suggested in ref. \citenum{zhang_exploring_2025} or more elaborate approaches in which a small set of descriptors is chosen by imposing that the on-sample fluctuations of $I^w(x)$  are minimal.  Such an avenue of progress will be explored in the near future

One of the features  which distinguishes our  approach from others  is the long-range properties of $V_{\mathcal{K}}(x)$, which can also be responsible for the success of the iterative approach for learning the committor in refs. \citenum{kang_computing_2024, trizio_everything_2025}
%Therefore, a natural evolution for this method would be to use graph neural networks \cite{zou_graph_2025, zhang_descriptor-free_2024}, which are direct functions of the atomic positions, and therefore should naturally encode all the relevant conformational changes.

Several avenues of further progress are possible.  The identification that a transition has taken place can be made automatically using methods like the one in ref. \citenum{faran_stochastic_2024} rather than having to resort to manual inspection.  Once a new state has been found, committor-based methods can be used to compute accurately the free energy surface and study the transition state ensemble.  In a metadynamics-like variant of the method, once the  $A\cup B$ has been explored, one could continue to look for other metastable states by iterating the procedure to speed-up further exploration. 
The use of an equivariant neural network will lift the need to use descriptors and start only with the Cartesian coordinates. 
This method is also ideally suited to generate on the fly new configurations on which machine learning potential can be trained. One can also envisage using this method as a way of directing the acquisition of new data in other machine learning applications. %\textcolor{red}{
This would amount to a reinforcement learning interpretation of what we do where  $V_{\mathcal K}(x)$  acts as a reward.
%}

Finally, it is our hope that this work brings the dream of using molecular dynamics as a totally blind exploration tool a bit closer.

%Finally, in this paper, we only discussed the part of the problem where we seek to discover possible next metastable states given an initial state. Once a new state is discovered, one can then extensively sample the reaction pathway using existing enhanced sampling tools such as the everything everywhere all at once recently developed committor-based approach  \cite{trizio_everything_2025} and determine both the free energy profile and the relative reaction rates.

\section{Methods}

A standard approach to enhanced sampling is based on filling the metastable states with an external bias that favors transitions to other states. In metadynamics and similar methods,  this is done by building on the fly the bias as a sum of repulsive Gaussians that are functions of a set of selected CVs \cite{laio_escaping_2002, invernizzi_exploration_2022, invernizzi_rethinking_2020}.  Even though ample literature has been devoted to the choice of  CV \cite{pietrucci_collective_2009,mouaffac_optimal_2023, karmakar_collective_2021,pietrucci_formamide_2015}, this step is often difficult, and an incorrect choice can affect the outcome. Furthermore, much wasted time is invested in reconstructing the free energy of the initial metastable state before basin $A$ is filled to a level high enough for the system to move out of $A$. 

In this paper instead, we construct a bias that encourages the system to sample new and yet unexplored regions, thus eventually inducing the system to leave the initial metastable state.  

In order to carry out this program, we first define a functional whose argument $I (x)$  is a function of the atomic coordinates, which is minimized by $I(x)=1$  if the configuration space is fully sampled. 
A simple example of one such functional is:
\begin{equation}
    \mathcal{L}[I]=\langle \vert \nabla_u I(x)\vert ^2\rangle  + \alpha (S-1)^2
\label{eq:simple_loss}
\end{equation}
  where the angular brackets indicate averages over the Boltzmann ensemble, $S=\langle I^2(x)\rangle$ is the second moment of $I(x)$,  $\nabla_u$  the mass-weighted gradients as in Kolmogorov functional, and the hyperparameter $\alpha$ controls the normalization of $I (x)$. 
However,  as an alternative to $\mathcal {L}\big([I])  $  we considered  using $\mathcal {E}\big([I], \lambda\big)  $  which depends on the function  $I(x)$  and the  scalar $\lambda$:
\begin{equation}
   \mathcal {E}\big([I], \lambda\big) = \dfrac{1}{(\eta +\lambda)^2} S(\eta S + M) - \dfrac{2}{\eta + \lambda}S + \alpha (1-S)^2 
   \label{eq:loss}
\end{equation}
where $S=\langle I^2(x)\rangle$ as before, $M = k_B T\langle |\nabla_u I(x)|^2 \rangle$ and $\eta$ a regularization parameter. This functional is minimal for $I(x)=1$ and $\lambda=0$ and is a limiting case of the one introduced in our earlier works\cite{devergne_biased_2024,devergne_slow_2025} (see SI). Since we found that $\mathcal{E}\big([I], \lambda\big)  $ has better numerical properties than $\mathcal{L}[I]$, $\mathcal {E}\big([I], \lambda\big)$ will be used in the following. 

%\textcolor{red}{ This might be due to the fact that when computing $\mathcal L$, the term that controls S is linearly added to the first $S$ independent term.   In contrast, in $\mathcal E$,   S also appears in all the terms of the functional.  }

In the practice, we shall transform the functional into a loss function and represent $I(x)$ as a neural network $I^w (x)$ where $w$ are variational parameters. These variational parameters will be optimized by minimizing the loss function through a stochastic gradient descent algorithm\cite{kingma_adam_2017}.

Of course, given an extensive sampling of the Boltzmann distribution, the solution of this variational problem is $I^w(x) = 1$ for all $x$. If however, only one basin, say $A$, has been sampled,  the solution will be $I^w(x) \approx 1$ for $x \in A$, but $I^w (x)\neq 1$ for $x\notin A$, since in this region no data are available and the neural network by necessity has to generalize $I^w (x)$. This leads to a change of slope as one crosses the boundary of the explored region.

In practice,  $I^w(x)$ is not perfectly constant, and thus, the Kolmogorov bias is not exactly zero in the training basin, which could lead to artificial barriers in the initial well. To overcome this problem, we complement the action of $V_{\mathcal K}(x)$ by an OPES Explore term, in a fashion similar to what is done in ref. \citenum{trizio_everything_2025},  and use  $I^w (x)$ as  collective variable.
%To accelerate even further the exploration, we proceed as in ref.  [\citenum{trizio_everything_2025}] and increase the algorithm  exploratory efficiency by adding to $V_{\mathcal K}(x)$  a bias that is built on the fly with \textsc{OPES} Explore\cite{invernizzi_exploration_2022}  using $I^\theta (x)$ as  collective variable. 
In addition,  the variational function that we insert  in equation \ref{eq:loss} has the form
$$I^w (x)= e^{-z^w (x)}.$$
Where $z^w$ is a neural network. The use of the exponential as activation function leads to a smoother $I^w (x)$ and guides the extrapolation.  Different activation functions could be devised, but this first choice has proven efficient enough for our purposes.

 When performing \textsc{OPES} simulations, one has to choose a \textsc{barrier} parameter \cite{noauthor_plumed_nodate}, which corresponds to the maximum barrier the system is expected to overcome during the simulation. In order to reduce the number of parameters to be set to a bare minimum, the value of $\lambda/\beta $ for $V_\mathcal{K}$ is chosen such that the minimum value of $V_{\mathcal{K}}(x_i)$ among the sampled $x_i$ is the \textsc{barrier} parameter.
  A low \textsc{opes barrier} value will force the system to follow a low free energy path, while a high value will allow exploring excited metastable states. 

Numerical details of the calculations will be presented in the  SI.  

\section{Acknowledgement
}%The  title is a  translation of  the words   " ...per montagne e per valloni.."  
We want to thank Enrico Trizio for helping implement the loss and the models in the open-source \textsc{mlcolvar} library.  We would like also to thank Peilin Kang for suggesting that the unusual properties of a   $V_{\mathcal K}(x)$-like  might be responsible for the recent success of our committor studies, where also a bias of a similar nature is used. We acknowledge that the research activity herein was carried out using the IIT Franklin HPC infrastructure; we gratefully acknowledge the Data Science and Computation Facility and its Support Team for their support and assistance on the IIT High Performance Computing Infrastructure.
This work was partially funded by the European Union - NextGenerationEU initiative and the Italian National Recovery and Resilience Plan (PNRR) from the Ministry of University and Research (MUR), under Project PE0000013 CUP J53C22003010006 "Future Artificial Intelligence Research (FAIR)".

\bibliographystyle{unsrt}
%\bibliography{bibliography.bib}
%\input{main.bbl}

\clearpage
\onecolumngrid
\setcounter{section}{0}         % Reset section counter
\renewcommand{\thesection}{S\arabic{section}}  % Number SI sections as 1, 2, 3...
\renewcommand{\thesubsection}{\thesection.\arabic{subsection}} % Optional: Subsections as 1.1, 1.2...

\vspace*{2em}
\begin{center}
    {\LARGE \textbf{Supporting Information}}
\end{center}
\vspace{1em}
\newcommand{\GE}{\mathcal L} %Infinitesimal generator

\newcommand{\state}{\mathbf{R}}

\date{June 2024}

%\begin{document}
\title{Supporting information}

%\maketitle

\section{Background on Langevin dynamics}
We assume the long-time behavior of the relaxation from an initial probability distribution towards equilibrium can be modeled by a Langevin equation. For a system of N atoms of mass ${m_i}$ and positions $\mathbf{R} = (\mathbf{R_1} \dots \mathbf{R_N}$), at temperature $T$ with friction parameter $\gamma$ this is given by:
\begin{equation}
    \frac{d\mathbf{R}_i}{dt} = -\frac{1}{\gamma m_i} \nabla U(\mathbf{R})) + \sqrt{\frac{2k_BT}{\gamma m_i}}\mathbf{\eta(t)}.
\end{equation}
Starting from an initial probability distribution $p_0(\mathbf{R})$, the system will relax towards its equilibrium probability distribution, which is the Boltzmann one given by $\pi(\mathbf{R})=e^{-\beta U(\mathbf{R})}/Z$ with $Z=\int e^{-\beta U(\mathbf{R})}d\mathbf{R}$ the partition function.

Instead of tracking the temporal evolution of the positions of individual atoms, we adopt a broader, more general vision by computing the time evolution of the associated probability distribution, $p_t(\mathbf{R})$, which is given by the Fokker-Planck equation. To ease the task of solving this equation, instead of directly computing the time evolution of the probability distribution, we compute that of its ratio with the Boltzmann distribution: $u_t=p_t/\pi$. The time evolution is given by the Backward Kolmogorov equation:

\begin{equation}
    \frac{\partial u_t}{\partial t}(\state) = -\frac{1}{\gamma}\sum_i^N \frac{1}{m_i}\frac{\partial u_t(\state)}{\partial r_i} \frac{\partial U(\mathbf{R})}{\partial r_i} + \frac{1}{\beta \gamma } \sum_i^N \frac{1}{m_i}\frac{\partial^2 u_t(\state)}{\partial r_i^2}.
    \label{eq:kolmogorov}
\end{equation}

This can be written in an operator form:

\begin{equation}
    \frac{\partial u_t}{\partial t}(\state) = -\mathcal{L}u_0
\end{equation}
where the action of $\GE$ on a twice differentiable function $f$ is:
\begin{equation}
    \GE f (\state) = \frac{1}{\gamma}\sum_i^N  \frac{1}{m_i}\frac{\partial f(\state)}{\partial r_i} \frac{\partial U(\mathbf{R})}{\partial r_i}  - \frac{1}{\beta} \sum_i^N \frac{1}{\gamma m_i}\frac{\partial^2 f(\state)}{\partial r_i^2}.
\end{equation}

Finally, we also recall that the matrix elements of $\GE$ between   two functions $\phi(\textbf{R})$ and $\psi(\textbf{R})$ can be computed  as
\begin{equation}
 \braket{\psi |\GE\vert \phi} = \int \frac{e^{-\beta U(\textbf{R})}}{Z} \psi(\textbf{R})^* \mathcal{L} \phi(\textbf{R}) d\textbf{R} = \frac{1}{\beta\gamma}\int \frac{e^{-\beta U(\textbf{R})}}{ Z} \nabla_{\textbf{u}} \psi(\textbf{R})^{*}\nabla_{\textbf{u}}\phi(\textbf{R}) d\textbf{R}
\end{equation}

\section{Learning $\GE$}
It can be shown \cite{devergne_biased_2024, devergne_slow_2025}, by using the resolvent $(\eta I -\GE)^{-1}$ of the infinitesimal generator, that the right eigenfunctions $\psi_i$ and eigenvalues $\lambda_i$ of $\GE$ minimize the following functional: 
\begin{equation}
    \label{eq:loss}
    E[\{ \psi_i \};\{\lambda_i\}]=\sum_{i=0}^m\sum_{j=0}^m S_{i,j} \frac  {1}{\eta +\lambda_j} W_{ij}\frac  {1}{\eta +\lambda_i}  
    -  2 \sum_i^m \frac  {1}{\eta +\lambda_i} S_{i,i}
\end{equation}

where $S$ is the overlap  matrix  whose entries are $S_{i,j}= \langle \psi_i \vert \psi_j \rangle $ and $W$ is a $(m+1) \times (m+1)$ matrix whose entries are: $W_{i,j}=\braket{\psi_j|(\eta I \!+\!\GE) |\psi_i}$.

The eigenpairs can thus be parametrized by neural networks with weights $w$ and optimized to minimize a loss which is the sum of the term in equation \ref{eq:loss} and an orthonormality term so that the total loss is:
\begin{equation}
\label{eq:dnn_loss_cov}
 E_\alpha(w, {\lambda_i}) = E[\{ \psi_i^w \};\{\lambda_i\}] + \alpha  \text{Tr}(S^w-1)^2.
\end{equation}

The first eigenfunction of $\GE$ corresponds to equilibrium, it is therefore constant, and it is clear that when $m=0$, only this function is learned, it will thus be constant over the training set. And therefore, the development of this paper applies to this loss. For $m=0$, the loss is simplified in:
\begin{equation}
    E_\alpha(w, \lambda) = \frac{1}{(\eta +\lambda)^2} S(\eta S + M) - \frac{2}{\eta + \lambda}S + \alpha (1-S)^2  
\end{equation}
where $S=\langle \psi^2\rangle$, $M = k_BT |\nabla \psi|^2\rangle$. It is thus now easy to see that this functional is minimized for $\psi=1$ and $\lambda=0$

\section{OPES simulations}
In most cases, the barriers to go from one metastable state to the other are too high compared to the thermal fluctuations. To overcome this problem, biased simulations have been set into place, where an additional term $V$ is added to the potential energy function $U$. One of the standard ways of building such a bias potential is by using On-the-fly Probability Enhanced Sampling. In this work, we use the Explore variation of this framework. The bias potential is given by:
%In this framework, one wishes to sample a target probability distribution $p_{tg}(s)$, where $s$ is the collective variable. To do so, a biased potential is introduced in the form of:
\begin{equation}
    V=\frac{1}{\beta}
    (\gamma-1)\log({p^{WT}_n(s)/Z_n+\epsilon})
\end{equation}
Where $p^{WT}_n = \frac{1}{n}\sum_{k=1}^nG(s,s_k)$ is the estimation at the $n$-th time step of the well-tempered probability distribution using kernel density estimation. $\epsilon$ is a regularization parameter. From this, the $\epsilon$ parameter, and the $\gamma$ parameter can be linked in the following way:
$\epsilon = e^{-\beta \Delta E/(1-1/\gamma)}$, where $\Delta E$ is the typical barrier of the system, which we called the \textsc{barrier} parameter in the main text.

\section{Alanine dipeptide}
\begin{figure}
    \centering
    \includegraphics[width=0.5\linewidth]{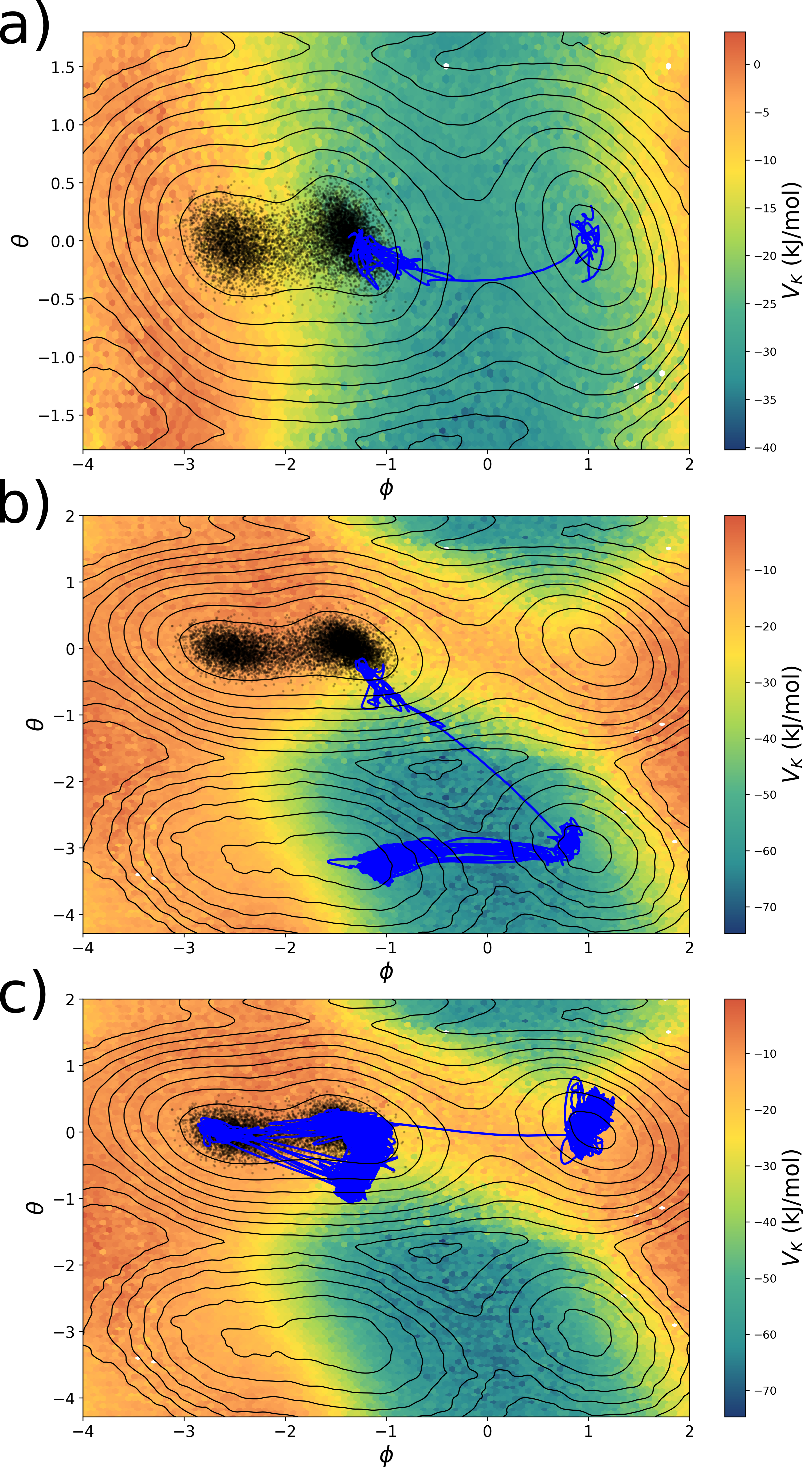}
    \caption{ 
     It can be seen that, in principle, four metastable states are possible. The lowest ones $C5, C7_{eq},\,\text{and}\, C7_{ax}$ are located in the upper part of the surface 
    Panel a): (blue) An example of an alanine dipeptide trajectory under the combined action of the Kolmogorov bias and of \textsc{OPES} explore, for a model that uses the interatomic distances as descriptors, which explores the lowest free energy minimum. 
    Panel b): (blue)  Same as a), but the simulation explores a higher free energy state. Panel c): same setup as panel b), but with a cutoff added to the Kolmogorov bias. In all panels, the training points are represented as black dots. 
    The level lines indicate the underlying free energy surface, and the background color represents the Kolmogorov bias. }
    \label{fig:SI_alanine}
\end{figure}

\textbf{Simulations details: }
All the simulations are run with GROMACS 2022.3 \cite{abraham_gromacs_2015} and patched with plumed 2.10 \cite{tribello_plumed_2014} in order to perform enhanced sampling simulations. We used the Amber99-SB \cite{salomon-ferrer_overview_2013} force field. The Langevin dynamics was sampled with a time step of 2fs with a damping coefficient $\gamma =1/0.05 ps^{-1}$ at a target temperature of 300K. For all the simulations, the \textsc{barrier} parameter of the \textsc{opes} Explore method was set to 40kJ/mol. 

The prefactor for the Kolmogorov bias was chosen so that its amplitude in the training set is  40 kJ/mol.

\textbf{Training with angles}
When looking for $I^{w}(x)$ as a function of the dihedral angles of the system, we used the sine and cosine of each angle for the input of the networks and an architecture of [8,20,20,1]. 

\textbf{Training with distances}

In the main text, we only presented the results where $I^{w}(x)$ is a function of only the four dihedral angles. However, such good variables are often not available. Here, we present the results when using the interatomic distances between heavy atoms as descriptors.

Like before, we trained 50 different models and ran each model for 5 ns by setting the \textsc{barrier} parameter again at  40kJ/mol. In this case, the total number of transitions is similar to the dihedral angle case. In fact, of the 50 trajectories 40 left the initial basin in the allotted time but only half of them ended in the expected final state while the others ended up in a higher energy conformational state (see figure \ref{fig:SI_alanine} panel c.) which differ from  C7ax   for the value of the  $\theta$ angle.  Interestingly, if we kept the simulation going, the system discovers yet another excited conformational state which is also in the lower $\theta$  quadrant.  This calculation is a first illustration of the ability of our method to discover new reactive pathways even in problems that, like alanine, have been thoroughly investigated.

On the other hand, one might want to limit the exploration behavior of the method to metastable states that are connected by the lowest free energy path. This can be done by lowering the \textsc{barier} parameter and adding a cutoff to the value of $V_{\mathcal{K}}$. We showcase this by taking the simulation in figure \ref{fig:SI_alanine}, panel c) and lowering the \textsc{barrier} parameter to 30kJ/mol. The resulting simulation is shown in figure \ref{fig:SI_alanine}, panel d).  We see that the system is first trying to explore high free energy regions, but it is restricted by the cutoff to remain in the $A$ basin until it transits to the  C7ax  state.

\textbf{Training details with distances: }

 We used the following parameters. For all the models, we chose the shift parameter $\eta=0.05$, and the orthonormality parameter $\alpha=50$. The training was performed using the ADAM optimizer for 20000 epochs with a learning rate of $5.10^{-4}$. We used an architecture of [45,32,32,1] node/layer for all the networks.

\textbf{Simulation with a cutoff}
When performing simulations to only explore low free energy states, we employed the Kolmogorov bias with a cutoff of $30$ kJ/mol, which was also the \textsc{opes barrier} parameter used.

\section{Claisen rearrangement}
\textbf{Simulation details: } For this system, we performed simulations with the CP2K code with the version 9.1.0 patched with plumed 2.9 at the PM6 level of theory. We set the integration timestep to 0.5fs. The NVT ensemble was sampled using the velocity rescaling thermostat with a temperature of 300 K and a time constant of 100 fs. For all simulations, we used a \textsc{barrier} parameter of 200 kJ/mol and a cutoff for the Kolmogorov bias at 400 kJ/mol

\textbf{Training details: }When looking for $I^w(x)$, as a function of all the interatomic distances between heavy atoms, we used the following parameters. For all the models, we chose the shift parameter $\eta=0.05$, and the orthonormality parameter $\alpha=50$. The training was performed using the ADAM optimizer for 20000 epochs with a learning rate of $5.10^{-4}$. We used an architecture of [45,32,32,1] node/layer for all the networks.

\section{alanine tetrapeptide}
\textbf{Computational details}
All the simulations are run with \textsc{gromacs 2022.3} \cite{ABRAHAM201519} and patched with plumed 2.10 \cite{TRIBELLO2014604} in order to perform enhanced sampling simulations. We used the Amber99-SB \cite{amber} force field. The Langevin dynamics was sampled with a timestep of 2fs with a damping coefficient $\gamma_ i =1/0.05 ps^{-1}$ at a target temperature of 300K. For all simulations, we used a \textsc{barrier} parameter of 80 kJ/mol

\textbf{Training details:} When looking for $I^w(x)$, as a function of all the 190 interatomic distances between heavy atoms, we used the following parameters. For all the models, we chose the shift parameter $\eta=0.05$, and the orthonormality parameter $\alpha=50$. The training was performed using the ADAM optimizer for 30000 epochs with a learning rate of $5.10^{-4}$. We used an architecture of [190,128,64,1] node/layer for all the networks.
\subsection{Additional simulations}
\begin{figure}
    \centering
    \includegraphics[scale=0.5]{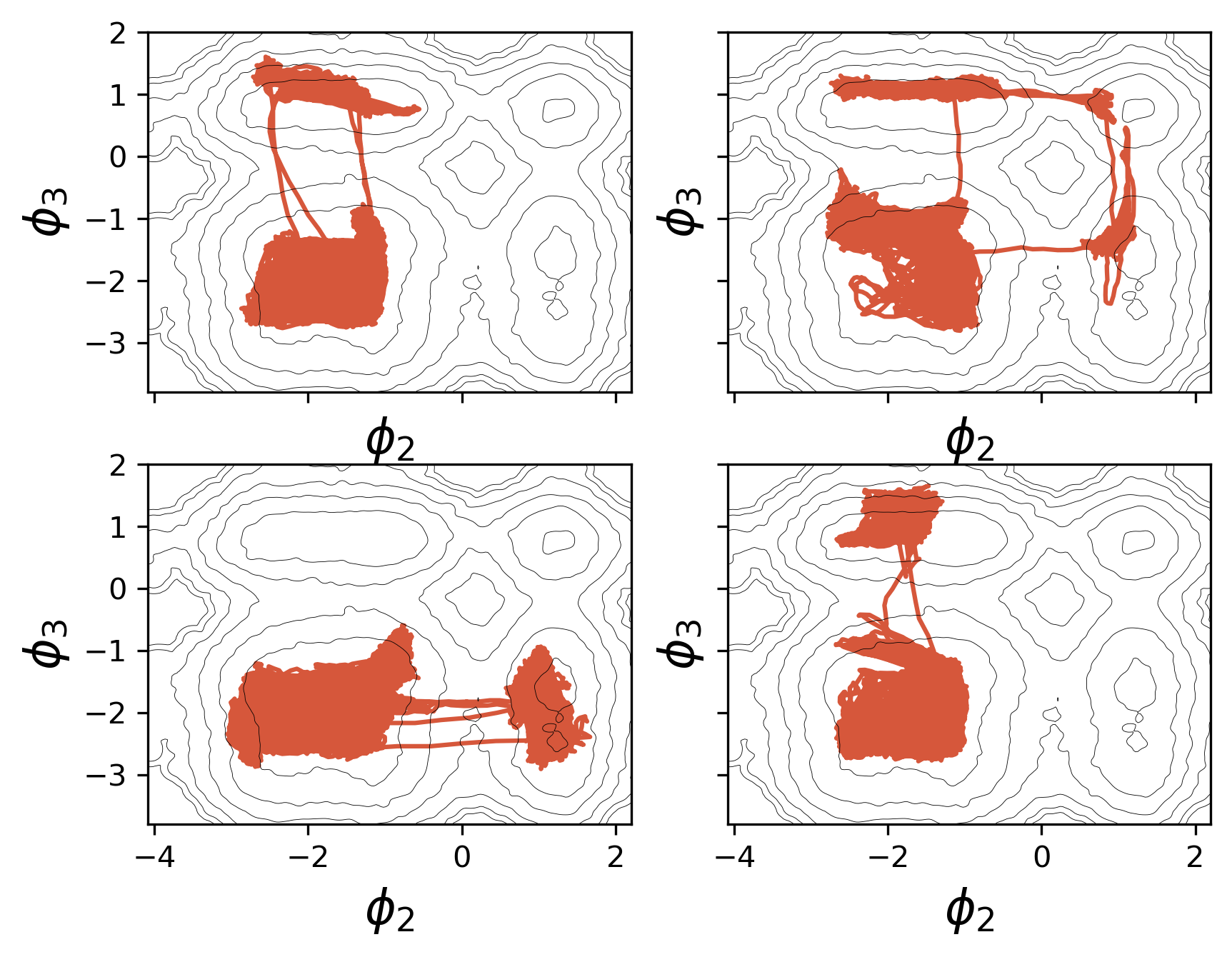}
    \caption{Results from four different simulations for the alanine tetrapeptide molecule. }
    \label{fig:si_tetrapeptide}
\end{figure}

Additional simulations are presented in figure \ref{fig:si_tetrapeptide}
\bibliographystyle{unsrt}
%\bibliography{bibliography}
%\input{bibliography_SI.bbl}
%\end{document}
\end{document}